# Combined Magnetic Imaging and Anisotropic Magnetoresistance Detection of Dipolar Skyrmions

*Jin Tang, Jialiang Jiang, Ning Wang, Yaodong Wu, Yihao Wang, Junbo Li, Y. Soh, Yimin Xiong, Lingyao Kong\*, Shouguo Wang, Mingliang Tian, and Haifeng Du\**

Jin Tang, Yimin Xiong, Lingyao Kong, Mingliang Tian

School of Physics and Optoelectronics Engineering Science, Anhui University, Hefei, 230601, China

E-mail: LyKong@ahu.edu.cn

Jin Tang, Jialiang Jiang, Ning Wang, Yaodong Wu, Yihao Wang, Junbo Li, Yimin Xiong, Mingliang Tian, Haifeng Du

Anhui Province Key Laboratory of Condensed Matter Physics at Extreme Conditions, High Magnetic Field Laboratory, HFIPS, Anhui, Chinese Academy of Sciences, Hefei, 230031, China

E-mail: duhf@hmfl.ac.cn

Y. Soh

Paul Scherrer Institute, 5232, Villigen, Switzerland

Shouguo Wang

School of Materials Science and Engineering, Anhui University, Hefei 230601, China

**Abstract**

Magnetic skyrmions are localized particle-like nontrivial swirls that are promising in building high-performance topological spintronic devices. The read-out functions in skyrmionic devices require the translation of magnetic skyrmions to electrical signals. Here, we report combined real-space magnetic imaging and anisotropic magnetoresistance studies on dipolar skyrmions. A single skyrmion chain and single skyrmion are observed using Lorentz transmission electron microscopy imaging. Meanwhile, the field, helicity, and skyrmion count dependence of anisotropic magnetoresistance of the $Fe_3Sn_2$ nanostructures are obtained simultaneously. Our results demonstrate that the anisotropic magnetoresistance of skyrmions is independent of the helicity and proportional to the skyrmion count. Our work could promote read-out operations in skyrmion-based spintronic devices.

## 1. Introduction

Magnetic skyrmions are potential information carriers that could be applied in future high-performance spintronic devices owing to their intriguing particle-like electromagnetic properties.[1-11] Reading operations of single bits in skyrmion-based devices can be achieved by translating magnetic signals of single skyrmions to electrical signals,[12] such as the anisotropic magnetoresistance (AMR),[13-19] Hall resistivity,[20-25] magnetoresistance in a magnetic tunneling junction,[26, 27] Nernst effect,[28-30] and tunneling magnetoresistance through an atomic tip.[31-33] Combined real-space magnetic imaging and electric-transport measurements are essential to be performed to understand detailed magnetic texture-related electrical signals, such as skyrmion count, skyrmion size, skyrmion polarization, and helicity dependence of electric-transport behavior. Maccariello *et al.*, Zeisslerl *et al.*, and Song *et al.* have combined Hall measurements and real-space magnetic imaging techniques of magnetic force microscopy or scanning transmission X-ray microscopy.[20-22] Scarioni *et al.* have combined anomalous Nernst effect measurements and magnetic force microscopy magnetic imaging.[28] The relative Hall and anomalous Nernst effect changes have been shown to be proportional to the skyrmion count and skyrmion size.[20-22, 28]

Planar magnetoresistance has been a fundamental indirect method to characterize the magnetization orientation due to the known AMR effects.[34, 35] AMR effects have been widely applied in field-sensitive sensors.[36, 37] Recently, anomalies in AMR signals have been also used to distinguish between the nucleation and annihilation of skyrmions indirectly.[13-18] However, in order to distinguish between different magnetic textures from these measured AMR signals, separate magnetic imaging on a different sample or numerical simulation was done.[13-18] A combined magnetic imaging and magnetoresistance detection of skyrmions on the same device is essential to explore skyrmion-related AMR effects directly.

Here, we report on a combined Lorentz-transmission electron microscopy (TEM) magnetic imaging and AMR detection of dipolar skyrmions in confined $Fe_3Sn_2$ nanostructures.

Lorentz-TEM images the in-plane magnetization with a nanometric-scale spatial resolution and has become one of the most powerful tools in imaging novel magnetic structures.[38] For example, Lorentz-TEM realized the first real-space observation of Bloch-type skyrmions,[39] antiskyrmions,[40, 41] skyrmion bundles,[42] and magnetic bobbers.[43] Here, with the assistance of in-situ TEM magnetic imaging, we show that the relative AMR changes are proportional to the skyrmion count and independent of the skyrmion helicity.

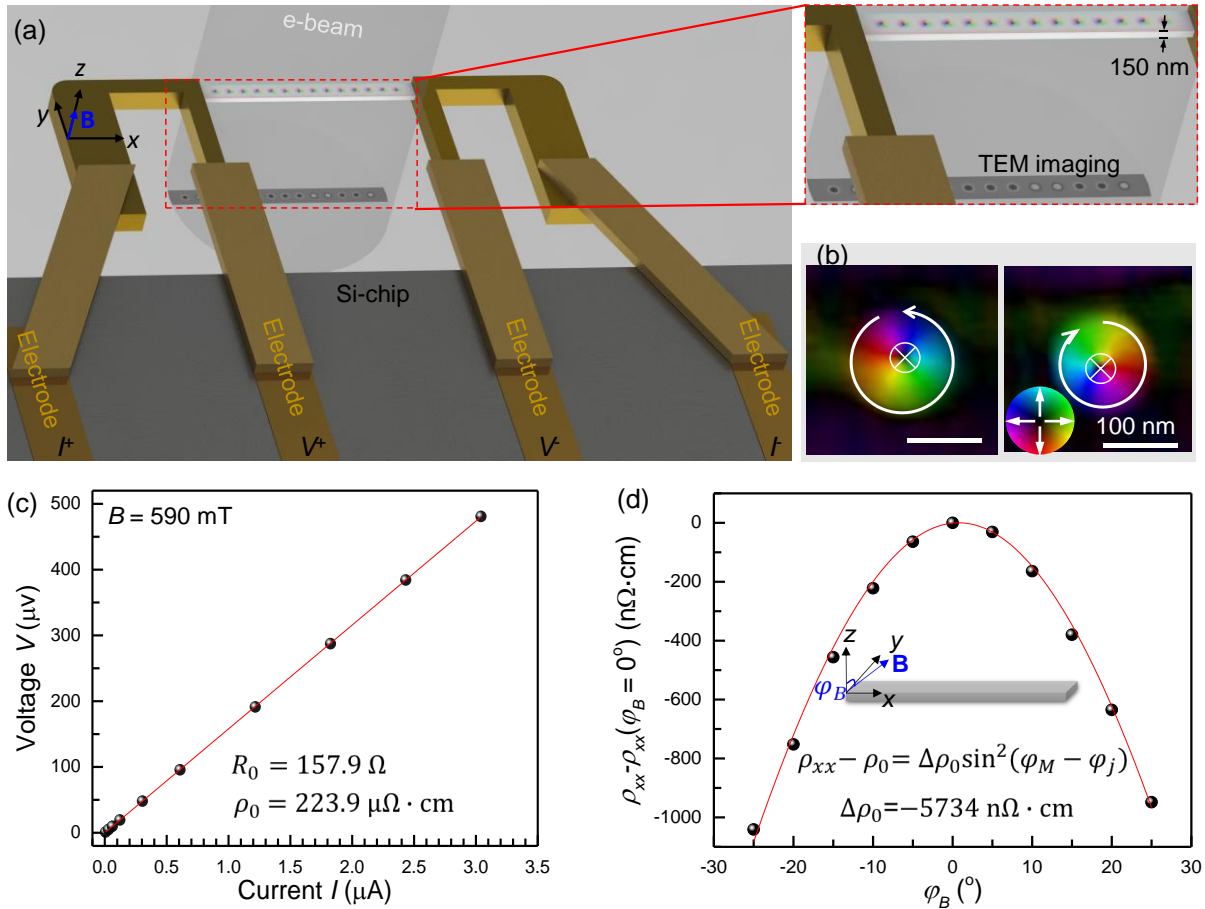

**Figure 1.** a) Schematic diagram for the combined TEM magnetic imaging and magnetoresistance detection of a single skyrmion chain. The width, length, and thickness of the device are 500 nm, 5500 nm, and 150 nm, respectively. b) Representative retrieved in-plane magnetization mapping of skyrmions with two helicities in the $Fe_3Sn_2$ nanostripe. The colors represent the in-plane magnetization amplitude and orientation based on the color wheel. c) Current dependence $I$ of voltage $V$ for the device at $B$ = 590 mT normal to the stripe

plane. $R_0$ is the resistance, $\rho_0$ is the resistivity. d) Tilting field angular dependence of magnetoresistance at $B$ = 1700 mT.

## 2. Results and Discussions

### 2.1. Combined studies of magnetic field dependence of magnetoresistance and magnetic texture

**Figure 1**a shows the experimental setup for the combined real-space observation and electrical detection of a single skyrmion chain in the $Fe_3Sn_2$(001) nanostripe. $Fe_3Sn_2$ is a kagome-lattice uniaxial ferromagnet with easy magnetization along the [001] axis at room temperature.[44-58] The dipolar skyrmions (also called skyrmion bubbles) in $Fe_3Sn_2$ with two helicities (Figure 1b) are stabilized by dipole-dipole interactions.[44-49] The confined $Fe_3Sn_2$ nanostripe with a 500 nm width enables the stabilization of a single skyrmion chain. We use the Fresnel magnetic imaging of Lorentz-TEM to obtain the magnetic evolution in the $Fe_3Sn_2$ nanostripe. The out-of-plane external magnetic field $B$ is provided by the objective lens of TEM. In addition, a standard four-terminal sensing method is used for the magnetoresistance measurements of the same device. From the linear current-voltage relation (Figure 1c), we obtain the device resistivity of ~223.9 $\mu\Omega \cdot cm$, which is consistent with that (~200 $\mu\Omega \cdot cm$) of bulk $Fe_3Sn_2$ single crystal,[50] and shows that the nanofabrication process does not degrade the sample quality. Our electronic diffraction pattern, high-resolution scanning transmission electron microscopy (HR-TEM), and energy dispersive spectroscopy (EDS) measurements (Supplemental Figure S1) also confirm that the chemical and structural orders of $Fe_3Sn_2$ microdevices are not destroyed in the nanofabrication process.

AMR effect describes the longitudinal resistivity dependent on the magnetization orientation with respect to the current flow direction. The AMR of a uniform ferromagnet (FM) is typically represented by:[34, 35]

$$\rho_{xx} = \rho_0 + \Delta\rho_0(\mathbf{u_I} \cdot \mathbf{u_m})^2 \qquad (1),$$

where $\rho_0$ is the magnetoresistance of the FM state at a saturation field, $\Delta\rho_0$ is the AMR coefficient, $\mathbf{u_I}$ is the unit vector along the current direction, and $\mathbf{u_m}$ is the unit vector along the magnetization direction. Here, $\mathbf{u_I}$ is along the *x* axis. The AMR coefficient $\Delta\rho_0$, which represents the difference in resistivity depending on the orientation of the magnetization with respect to the current, is obtained from fitting the tilted field orientation dependence of magnetoresistance (Figure 1d) at a constant field of ~1700 mT in the FM state. The magnetization orientation angle $\varphi_m$ can be taken to be parallel to the field orientation $\varphi_B$ since Fe$_3$Sn$_2$ is a soft FM that is saturated at a field of 1700 mT.[50] Here, $\varphi_m$ and $\varphi_B$ are the angle of the magnetization and magnetic field with respect to the out-of-plane *z* axis, respectively. According to the AMR effect expressed by Equation (1), the $\varphi_B$-$\Delta\rho_{xx}$ relation is fitted by a function: $\Delta\rho_{xx} = \rho_{xx} - \rho_0 = \Delta\rho_0 \left[\cos\left(\frac{\pi}{2} - \varphi_B\right)\right]^2 = \Delta\rho_0 \sin(\varphi_B)^2$. We thus obtain $\Delta\rho_0 = -5734$ n$\Omega \cdot$ cm from the fitting (Figure 1d). The negative value of $\Delta\rho_0$ means that the resistivity is smaller when the magnetization is in the plane (anti)parallel to the current than out-of-plane perpendicular to the current. The sign and the magnitude of $\frac{\Delta\rho_0}{\rho_0} \approx$ 2.6 % are both consistent with the measured magnetoresistance change in bulk single-crystal Fe$_3$Sn$_2$ further supporting that the nanofabrication process does not degrade the sample quality.[50]

We first explore the magnetoresistance and magnetic evolution in the Fe$_3$Sn$_2$ nanostripe by sweeping the field from 0 to 600 mT, and then from 600 to 0 mT, as shown in **Figure 2**a. The magnetic field is applied perpendicular to the kagome plane, *i.e.*, $\varphi_B = 0°$. The magnetoresistance increases as the external field increases and reaches the maximum at *B* = 590 mT. The magnetoresistance decreases as the external field is turned to 0 mT. The magnetoresistance is independent of the field sign (inset of Figure 2a) and shows no hysteresis in the low-field region of ~0-460 mT. However, in the high-field region of ~460-600 mT, we observe a resistivity hysteresis with a lower resistivity value in the field increasing process. Real-space magnetic imaging on the same device verifies a magnetic domain hysteresis in such a high field region, where the skyrmion chain and FM coexist (Figure 2b and 2c). At zero field, stripe domains are the magnetic ground states (Figure S1). When the out-of-plane field increases, the stripes shrink and transform to skyrmions. A single skyrmion chain with skyrmion count $N_S$ = 14 is obtained at $B$ = 425 mT in the field increasing process (Figure 2b). We note the presence of skyrmions with different sizes, which can be attributed to width differences in the stripe (Figure S1). Skyrmions get larger in the nanostripe with larger widths at the same magnetic field (Figure S2). In the field range from 425 to 500 mT, the skyrmions shrink but the skyrmion count stays at 14 (Figure 2b). The skyrmions in the nanostripe annihilate one by one in the field range from 500 to 560 mT (Figure S3). A uniform FM is achieved at $B$ = 580 mT. In the field decreasing process, we observe a magnetic hysteresis in the field range from 600 to 465 mT (Figure 2c), where FM is observed. Thus, in the hysteresis field range between 465 and 600 mT, the skyrmion chain and FM are the preferred phases in the increasing and decreasing field processes, respectively. It is to be noted that the magnetic field of the Lorentz-TEM is always continuously changed in a few minutes between two measured fields. The magnetoresistance hysteresis suggests that the skyrmion chain and FM in the nanostripe can be electrically distinguished, with the magnetoresistance of the skyrmion state being smaller than that of FM. A lower magnetoresistance for the skyrmion state compared to the FM state is expected since a uniform FM state magnetized perpendicular to the kagome plane yields a higher resistivity than a uniform FM state magnetized along the current orientation in the kagome plane (Figure 1d). A skyrmion has a magnetic structure deviating from a uniform out-of-plane FM state and thus should exhibit a lower resistivity.[2, 5]

We micromagnetically simulate the corresponding magnetic field dependence of magnetic domain evolution and total free energy based on measured magnetic and geometrical parameters (Figure S4 and S5). The magnetization hysteresis in the high field region is also identified in our simulation. The skyrmion and FM states are both equilibrium states in a field range between 350 and 475 mT, with the skyrmion and FM being more stable states for $B < 425$ mT and $B > 425$ mT, respectively. The coexistence of the skyrmion chain and FM suggests an energy barrier between them. We further demonstrate the relative energy profiles in the increasing and decreasing $B$ processes (Figure S5c and S5d), which fits the magnetic hysteresis in our experiments.

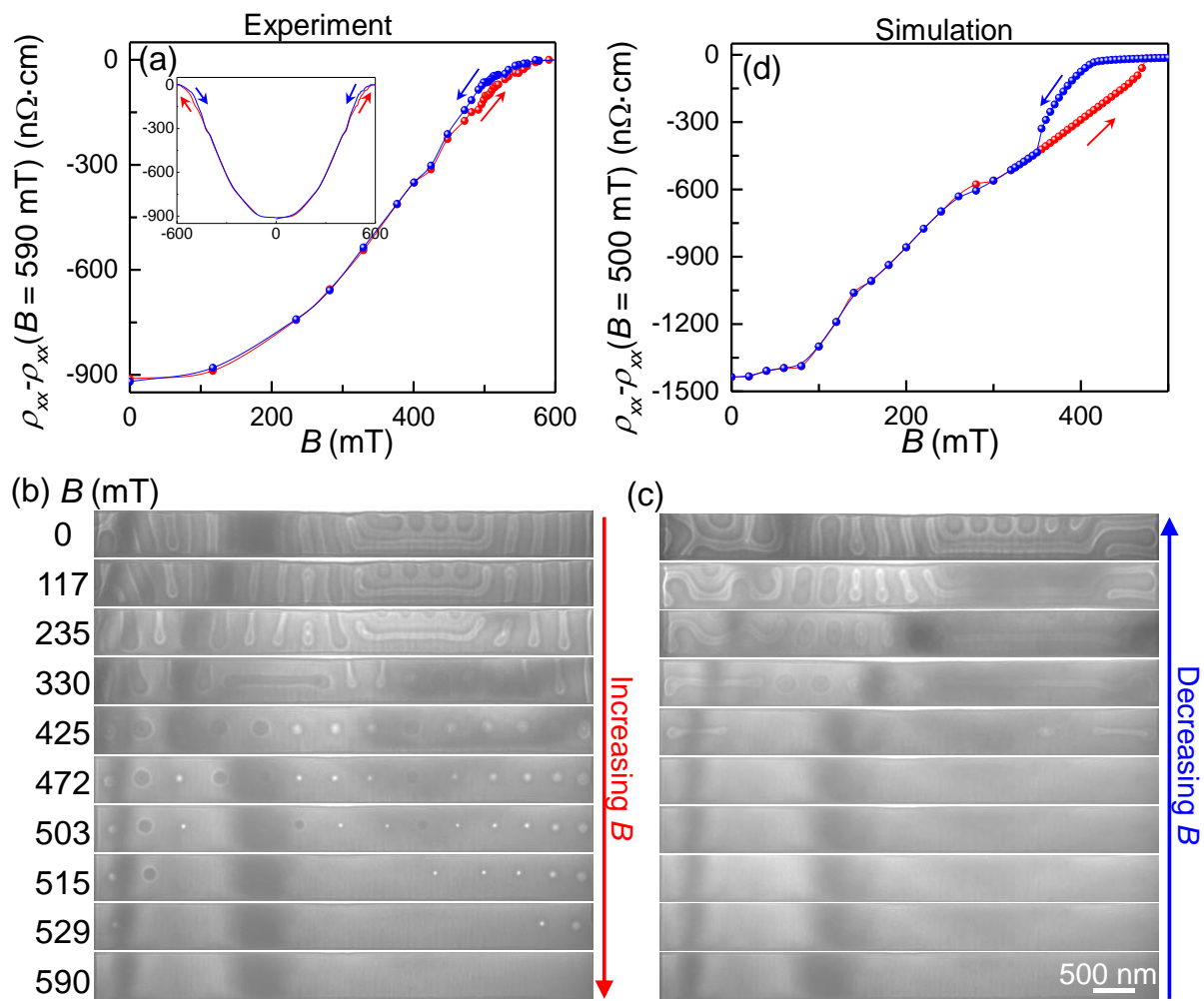

**Figure 2.** Field dependence of magnetoresistance. a) Magnetic field $B$ dependence of magnetoresistance in the field range from 0 to 600 mT. The inset shows magnetoresistance in the full field range from −600 to 600 mT. b) and c) Magnetic field $B$ dependence of magnetic

evolutions for increasing and decreasing field processes. Defocused distance −1000 μm. (d) Numerically calculated magnetic field dependence of magnetoresistance. $\varphi_B = 0°$.

For a non-uniform magnetization state, we assume that the magnetoresistance is the average AMR of each spin $\mathbf{m}_{i,j}$ and expressed by:

$$\rho_{xx} = \rho_0 + \Delta\rho_0 < \left(\mathbf{u}_\mathbf{I} \cdot \mathbf{u}_\mathbf{m}^{i,j}\right)^2 > \qquad (2)$$

We thus obtain the simulated field dependence of the magnetoresistance according to Equation (2) and simulated magnetic configurations (Figure S4), as shown in Figure 2(d). The simulated magnetoresistance reveals a comparable change (around 1400 nΩ·cm) and a similar trend as that of the experiment in the whole magnetic field region. In the high field region, the magnetic domain hysteresis (Figure S4) also contributes to the resistivity hysteresis in the simulation (Figure 2d). Simulation shows that the FM state reveals a non-uniform magnetization in the near-surface layers (Figure S6), which explains the resistivity change in the decreasing field process from $B$ = 600 mT to $B$ = 427 mT even though the magnetic state remains FM (Figure 2d).

## 2.2. Skyrmion helicity and count dependence of magnetoresistance

Since skyrmions (also called skyrmion bubbles) in the centrosymmetric $Fe_3Sn_2$ are stabilized without chiral interactions, the skyrmions with two helicities at a given out-of-plane field are in energy equilibrium. Thus, counter-clockwise and clockwise rotations of skyrmions with identical polarizations can be observed simultaneously in the nanostripe (Figure S1). We explore the skyrmion helicity dependence of magnetoresistance in the $Fe_3Sn_2$ nanostripe, as shown in **Figure 3**a. We fix the field at 490 mT. When repeating the process of increasing the field from zero to $B$ = 490 mT, we obtain 14 skyrmions at most but with random helicities, as shown in Figure 3b. The FM state at $B$ = 490 mT is obtained by decreasing the field from 600 mT. In comparison to the FM state, a relative resistivity change of single skyrmion chains with $N_S = 14$ as a function of skyrmion helicity is shown in Figure 3a. The relative resistivity change is around $-50$ nΩ·cm and shows no obvious trend with respect to the count of skyrmions with counterclockwise ($N_{S\text{-CCW}}$) or clockwise ($N_{S\text{-CW}}$) rotations. We also explore the simulated skyrmion helicity dependence of the AMR effect (Figure 3a). Because the skyrmions with clockwise and counter-clockwise rotations contribute the same average AMR effects according to Equation (2), the magnetoresistance of skyrmions is independent of helicity. Our simulation results verify that the AMR of the single skyrmion chain is independent of the helicity of the skyrmions. Simulated results show that the 14 skyrmions in the chain contribute $-100$ nΩ·cm of AMR effect, which is of the same order of magnitude as that ($\sim-50$ nΩ·cm) in our experiments.

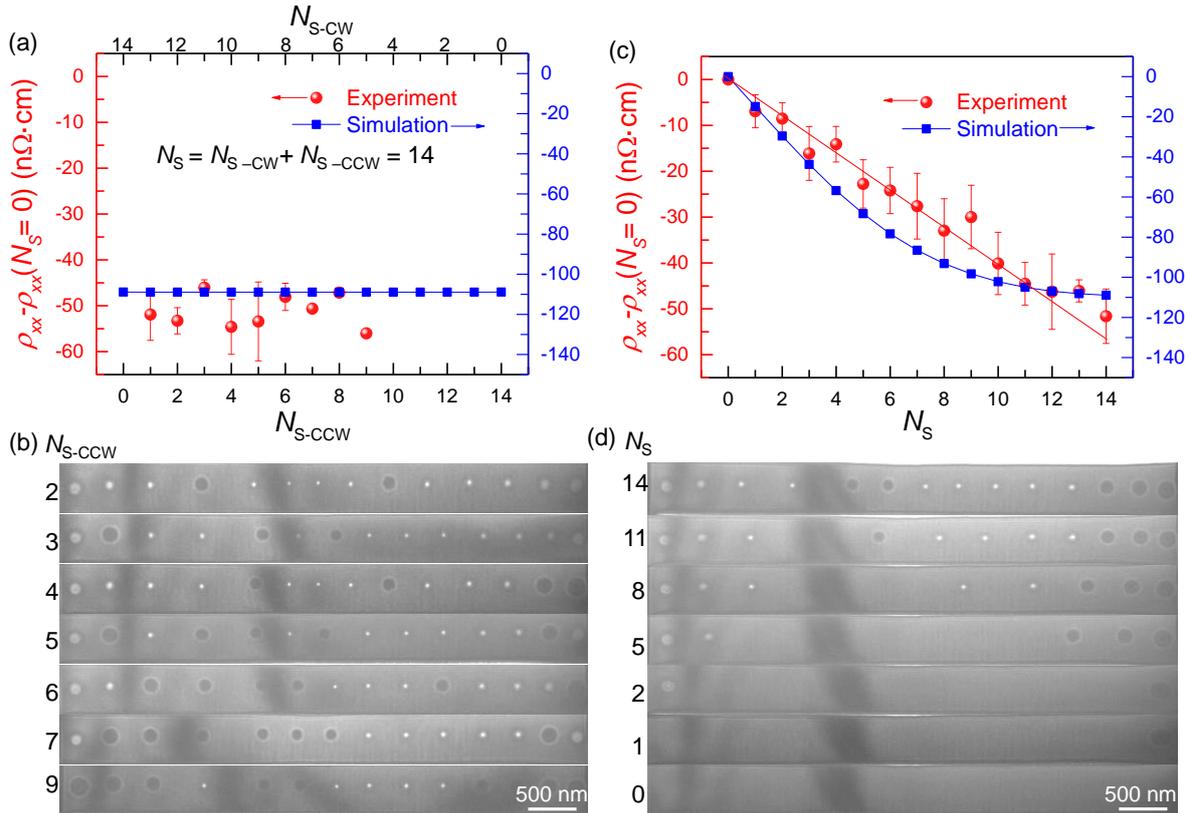

**Figure 3.** Skyrmion helicity and count dependence of magnetoresistance. a) Dependence of magnetoresistance at a fixed field on the count of skyrmions with counter-clockwise rotation $N_{S\text{-CCW}}$ or clockwise rotation $N_{S\text{-CW}}$. Experiment $B = 490$ mT and simulation $B = 455$ mT. b) Fresnel magnetic images of a single skyrmion chain with different helicities at $B = 490$ mT. The total skyrmion count $N_S$ of the single skyrmion chain is fixed at 14. c) Skyrmion count

dependence of magnetoresistance at a fixed field. Experiment $B = 490$ mT and simulation $B = 455$ mT. d) Fresnel magnetic images of a single skyrmion chain with varied skyrmion count at $B = 490$ mT. The error bars in (a) and (c) denote the standard deviation obtained from multiple measurements. The total individual measurements in (a) are 2 for $N_{s\text{-CCW}} = 1$, 2 for $N_{s\text{-CCW}} = 2$, 3 for $N_{s\text{-CCW}} = 3$, 3 for $N_{s\text{-CCW}} = 4$, 4 for $N_{s\text{-CCW}} = 5$, 4 for $N_{s\text{-CCW}} = 6$, 1 for $N_{s\text{-CCW}} = 7$, 1 for $N_{s\text{-CCW}} = 8$, 1 for $N_{s\text{-CCW}} = 9$. The total individual measurements in (c) are 19 for $N_s = 14$, 8 for $N_s = 13$, 9 for $N_s = 12$, 7 for $N_s = 11$, 4 for $N_s = 10$, 10 for $N_s = 9$, 6 for $N_s = 8$, 8 for $N_s = 7$, 5 for $N_s = 6$, 10 for $N_s = 5$, 7 for $N_s = 4$, 6 for $N_s = 3$, 3 for $N_s = 2$, 10 for $N_s = 1$. In individual measurements, the state of the skyrmion chain has been changed by cycling the magnetic field. Defocused distance $-1000$ μm. $\varphi_B = 0°$.

We have shown a magnetoresistance change of $-50$ nΩ·cm between a single skyrmion chain with $N_S = 14$ and FM with $N_S = 0$. We further explore the skyrmion count dependence of AMR in the $Fe_3Sn_2$ nanostripe at a fixed field of $B = 490$ mT. Because skyrmions are annihilated one by one in the field region of 500-600 mT, we are able to obtain single skyrmion chains with different skyrmion counts at higher fields. The skyrmion counts in single skyrmion chains stay fixed when decreasing the field back to 490 mT because of magnetic hysteresis (Figure S3). We measure the relative magnetoresistance change of single skyrmion chains with tunable skyrmion counts, as shown in Figure 3c and 3d. The magnetoresistance change increases almost linearly as the skyrmion counts decrease to 0. In the $Fe_3Sn_2$ nanostripe with a length of ~5500 nm, the relative magnetoresistance change contributed by a single skyrmion is about $-3.86$ nΩ·cm. We also explore the skyrmion count dependence of AMR by simulations (Figure 3c). When the skyrmion count decreases, the AMR increases as expected. However, the AMR is not a linear function of the skyrmion count in the simulation. The non-linear dependence of the AMR on the skyrmion count in the simulation is attributed to the increased skyrmion size for decreased skyrmion count in the chain (Figure S7).

## 2.3. Detecting a single skyrmion

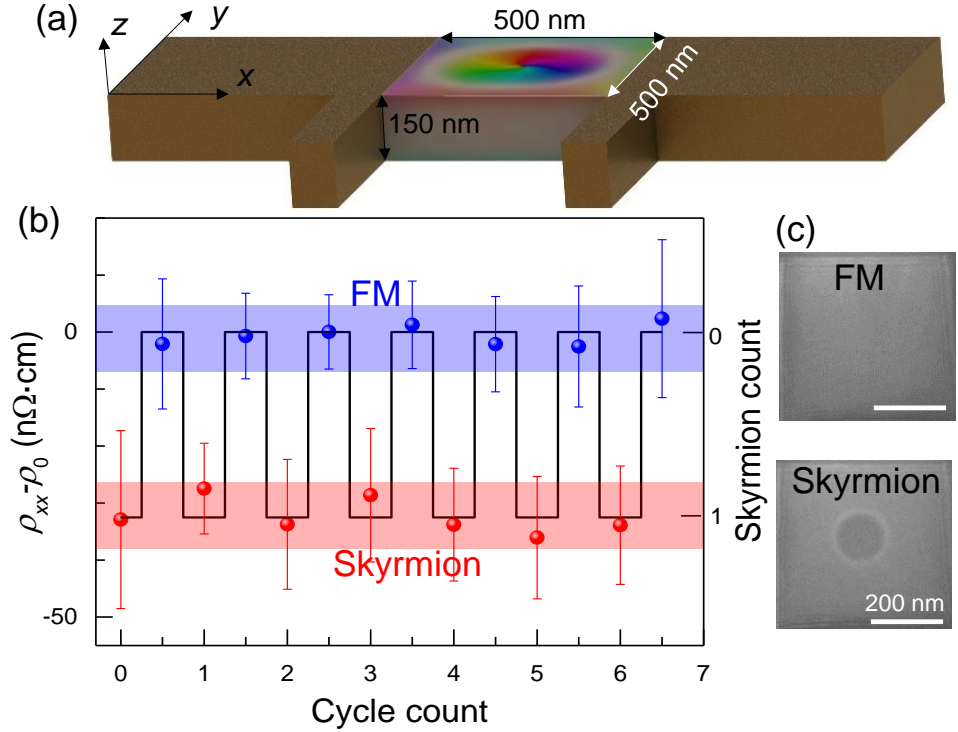

**Figure 4.** a) Schematic of the device for detecting a single skyrmion in a confined $Fe_3Sn_2$ cuboid by both magnetoresistance and magnetic imaging. b) The variation of magnetoresistance in the presence and absence of the skyrmion. The error bar is the standard deviation of ~50 iterations in each individual measurement. c) The associated defocused Fresnel magnetic imaging of the skyrmion and FM. Defocused distance −1000 μm. $\varphi_B = 0°$.

We finally fabricate a strongly confined $Fe_3Sn_2$ cuboid with a width and length of ~500 nm and thickness of ~150 nm, as shown in **Figure 4**a. In the $Fe_3Sn_2$ cuboid, one dipolar skyrmion at most can be stabilized. The skyrmion state is achieved by increasing $B$ from 0 to 540 mT and the FM state is achieved by decreasing $B$ from 600 to 540 mT (Figure 4c). The magnetoresistance $\rho_{xx}$ at a fixed field of 540 mT in the presence or absence of a single skyrmion is obtained with a difference of ~−35 nΩ · cm (Figure 4c), with the skyrmion state yielding a lower resistivity state. Based on Equation (2) and the micromagnetic simulated skyrmion and FM states in the confined cuboid, we obtain the theoretically maximum magnetoresistance change for Figure 4a contributed by a single skyrmion to be −38 nΩ · cm, which is comparable to our experimental result (−35 nΩ · cm). Note that the value of the resistivity $\rho$ change between the FM state and the skyrmion state in the case for a cuboid (−35 nΩ · cm) is comparable to that between the FM state and the skyrmion chain state with $N_S = 14$ in the case of a nanostripe (~−50 nΩ · cm). This is due to the fact that both devices can be regarded as the same electrical medium except for their length (assuming that the skyrmion linear density is the same in both devices) and resistivity is a quantity that is independent of the length. To make the skyrmion linear density the same in both devices, the correct skyrmion chain state to compare to the cuboid is one with $N_S = 11$ (~−42 nΩ · cm) since the length of the stripe is 11 times that of the cuboid. Given that the AMR effect is roughly linear to $N_S$, this explains why the AMR effect from a single skyrmion in our nanostripe (−3.86 nΩ · cm) is about one order of magnitude smaller than in our cuboid when expressed in resistivity.

## 3. Conclusions

In summary, we have demonstrated the electrical detection of dipolar-stabilized skyrmions in the $Fe_3Sn_2$ nanostructures with the assistance of in-situ real-space TEM magnetic imaging. The combined magnetic imaging and AMR detection enable the determination of magnetotransport behavior dependent on detailed skyrmion properties: AMR effect cannot distinguish the skyrmions with different helicities; AMR can detect a single skyrmion chain with the amplitude determined by the skyrmion count; AMR can detect the presence of a single skyrmion. Because the skyrmion helicities and skyrmion counts are important degrees of freedom that could be used for information operation and storage,[5, 48, 59] our results could provide a guide in the further development of skyrmionic devices.

## 4. Methods

*Preparation of bulk Fe$_3$Sn$_2$ crystal*:[44, 45] Single Fe$_3$Sn$_2$ crystals were grown by chemical vapor transport with stoichiometric iron (Alfa Aesar, >99.9%) and tin (Alfa Aesar, >99.9%). The sintered Fe$_3$Sn$_2$ was obtained by heating the mixture at 800 ℃ for 7 days, followed by thorough grinding. It was then sealed with I$_2$ in a quartz tube under vacuum and kept in a temperature gradient of 720 ℃ to 650 ℃ for 2 weeks.

*Fabrication of Fe$_3$Sn$_2$ microdevices*: The 150-nm thick Fe$_3$Sn$_2$ nanostructured stripe and cuboid with standard four electrical contacts for resistivity measurements were fabricated from a bulk single crystal using a standard lift-out method, with a focused ion beam and scanning electron microscopy dual beam system (Helios Nanolab 600i, FEI).

*TEM measurements*: We used in-situ Fresnel imaging in Lorentz-TEM (Talos F200X, FEI) with an acceleration voltage of 200 kV to investigate magnetic domains in the Fe$_3$Sn$_2$ nanostructure. The TEM holder (model 613.6, Gatan) can support magnetoresistance measurements. To detect a small change in the AMR signal, we measured the resistance using the standard lock-in technique,[60] which reduces noise in the measurement. In the lock-in technique, the input current $I$ is not a direct current, but a sine wave expressed by $I = I_0\sin(2\pi ft)$, where $I_0$ is the amplitude, $f$ is the frequency, and $t$ is the time. The voltage $V$ between the two ends of devices (supplemental Figure S1b) was fed into the signal input channel of the lock-in amplifier. The magnetoresistance was measured using the lock-in technique (SR830) at a frequency of 317 Hz. All experiments were performed at 295 K.

*Micromagnetic simulations*: The zero-temperature micromagnetic simulations were performed using MuMax3.[61] We consider in the Hamiltonian exchange interaction ($A$) energy, uniaxial magnetic anisotropy ($K_u$) energy, Zeeman energy, and dipole-dipole interaction energy.[61] Simulated magnetic parameters are set based on the material $Fe_3Sn_2$ at room temperature ($K_u$ = 54.5 kJ·m$^{-3}$, saturated magnetization $M_s$ = 622.7 kA·m$^{-1}$, and $A$ = 8.25 pJ·m$^{-1}$) [44, 45]. The cell size was set at $2 \times 2 \times 3$ nm$^3$.

*Statistical Analysis:* We have performed multiple individual measurements to obtain the skyrmion count and helicity dependence of the AMR effect at a fixed field. In individual measurements, the state of the skyrmion chain was changed by cycling the magnetic field. The experimental resistivity denotes the mean value of multiple individual measurements. The error bar equals the standard deviation of multiple individual measurements.

**Supporting Information**

Supporting Information is available from the Wiley Online Library or from the author.

**Acknowledgments**

H. D. acknowledges the financial support from China's National Key R&D Program, Grant No. 2017YFA0303201; the Major/Innovative Program of Development Foundation of Hefei Center for Physical Science and Technology, Grant No. 2016FXCX001; the Strategic Priority Research Program of Chinese Academy of Sciences, Grant No. XDB33030100; the Youth Innovation Promotion Association CAS No. 2015267; and the Equipment Development Project of the Chinese Academy of Sciences, Grant No. YJKYYQ20180012. J. T., Y. X., Y. Wu, and L.K. acknowledge the financial support of the Natural Science Foundation of China, Grant Nos. 12174396, 12104123, 11974021, and U1432138. Y. X. acknowledges the financial support of China's National Key R&D Program, Grant No. 2016YFA0300404, and the Collaborative Innovation Program of Hefei Science Center, CAS, Grant No. 2019HSC-CIP007.

**Author Contributions**

H.D. and J.T. supervised the project and conceived the experiments. Y. Wang, J.L., and Y.X. synthesized the $Fe_3Sn_2$ bulk single crystals. J.T. performed the $Fe_3Sn_2$ microdevice fabrication, TEM, and magnetoresistance measurements with the help of J.J., Y. Wu, and N.W.. J.T. performed the simulations with the help of L.K.. J.T. and H.D. wrote the manuscript with input from all authors. All authors discussed the results and contributed to the manuscript.

**Conflict of Interest:**

The authors declare no competing financial interest.

# References

[1] A. Bogdanov and A. Hubert, *J. Magn. Magn. Mater.* **1994**, *138*, 255.

[2] S. Mühlbauer, B. Binz, F. Jonietz, C. Pfleiderer, A. Rosch, A. Neubauer, R. Georgii, and P. Böni, *Science* **2009**, *323*, 915.

[3] W.-S. Wei, Z.-D. He, Z. Qu, and H.-F. Du, *Rare Met.* **2021**, *40*, 3076.

[4] Y. Zhou, *Natl. Sci. Rev.* **2019**, *6*, 210.

[5] N. Nagaosa and Y. Tokura, *Nat. Nanotechnol.* **2013**, *8*, 899.

[6] Y. Tokura and N. Kanazawa, *Chem. Rev.* **2021**, *121*, 2857.

[7] B. Göbel, I. Mertig, and O. A. Tretiakov, *Phys. Rep.* **2021**, *895*, 1.

[8] K. Everschor-Sitte, J. Masell, R. M. Reeve, and M. Kläui, *J. Appl. Phys.* **2018**, *124*, 240901.

[9] C. H. Marrows and K. Zeissler, *Appl. Phys. Lett.* **2021**, *119*, 250502.

[10] W. Kang, Y. Huang, X. Zhang, Y. Zhou, and W. Zhao, *Proc. IEEE* **2016**, *104*, 2040.

[11]  A. Fert, N. Reyren, and V. Cros, *Nat. Rev. Mater.* **2017**, *2*, 17031.

[12]  S. S. Wang, J. Tang, W. W. Wang, L. Y. Kong, M. L. Tian, and H. F. Du, *J. Low Temp. Phys.* **2019**, *197*, 321.

[13]  H. Du, J. P. DeGrave, F. Xue, D. Liang, W. Ning, J. Yang, M. Tian, Y. Zhang, and S. Jin, *Nano Lett.* **2014**, *14*, 2026.

[14]  H. Du, D. Liang, C. Jin, L. Kong, M. J. Stolt, W. Ning, J. Yang, Y. Xing, J. Wang, R. Che, J. Zang, S. Jin, Y. Zhang, and M. Tian, *Nat. Commun.* **2015**, *6*, 7637.

[15]  M. J. Stolt, S. Schneider, N. Mathur, M. J. Shearer, B. Rellinghaus, K. Nielsch, and S. Jin, *Adv. Funct. Mater.* **2019**, *29*, 1805418.

[16]  N. Mathur, F. S. Yasin, M. J. Stolt, T. Nagai, K. Kimoto, H. Du, M. Tian, Y. Tokura, X. Yu, and S. Jin, *Adv. Funct. Mater.* **2021**, *31*, 2008521.

[17]  S. X. Huang, F. Chen, J. Kang, J. Zang, G. J. Shu, F. C. Chou, and C. L. Chien, *New J. Phys.* **2016**, *18*, 065010.

[18]  D. A. Gilbert, B. B. Maranville, A. L. Balk, B. J. Kirby, P. Fischer, D. T. Pierce, J. Unguris, J. A. Borchers, and K. Liu, *Nat. Commun.* **2015**, *6*, 8462.

[19]  T. Haug, A. Vogl, J. Zweck, and C. H. Back, *Appl. Phys. Lett.* **2006**, *88*, 082506.

[20]  D. Maccariello, W. Legrand, N. Reyren, K. Garcia, K. Bouzehouane, S. Collin, V. Cros, and A. Fert, *Nat. Nanotechnol.* **2018**, *13*, 233.

[21]  K. Zeissler, S. Finizio, K. Shahbazi, J. Massey, F. A. Ma'Mari, D. M. Bracher, A. Kleibert, M. C. Rosamond, E. H. Linfield, T. A. Moore, J. Raabe, G. Burnell, and C. H. Marrows, *Nat. Nanotechnol.* **2018**, *13*, 1161.

[22]  K. M. Song, J.-S. Jeong, B. Pan, X. Zhang, J. Xia, S. Cha, T.-E. Park, K. Kim, S. Finizio, J. Raabe, J. Chang, Y. Zhou, W. Zhao, W. Kang, H. Ju, and S. Woo, *Nature Electronics* **2020**, *3*, 148.

[23]  K. Hamamoto, M. Ezawa, and N. Nagaosa, *Appl. Phys. Lett.* **2016**, *108*, 112401.

[24]  P. Bruno, V. K. Dugaev, and M. Taillefumier, *Phys. Rev. Lett.* **2004**, *93*, 096806.

[25] A. Neubauer, C. Pfleiderer, B. Binz, A. Rosch, R. Ritz, P. G. Niklowitz, and P. Boni, *Phys. Rev. Lett.* **2009**, *102*, 186602.

[26] N. E. Penthorn, X. Hao, Z. Wang, Y. Huai, and H. W. Jiang, *Phys. Rev. Lett.* **2019**, *122*, 257201.

[27] S. Li, A. Du, Y. Wang, X. Wang, X. Zhang, H. Cheng, W. Cai, S. Lu, K. Cao, B. Pan, N. Lei, W. Kang, J. Liu, A. Fert, Z. Hou, and W. Zhao, *Science Bulletin* **2022**, *67*, 691.

[28] A. Fernandez Scarioni, C. Barton, H. Corte-Leon, S. Sievers, X. Hu, F. Ajejas, W. Legrand, N. Reyren, V. Cros, O. Kazakova, and H. W. Schumacher, *Phys. Rev. Lett.* **2021**, *126*, 077202.

[29] Z. D. Wang, M. H. Guo, H. A. Zhou, L. Zhao, T. Xu, R. Tomasello, H. Bai, Y. Q. Dong, S. G. Je, W. L. Chao, H. S. Han, S. Lee, K. S. Lee, Y. Y. Yao, W. Han, C. Song, H. Q. Wu, M. Carpentieri, G. Finocchio, M. Y. Im, S. Z. Lin, and W. J. Jiang, *Nature Electronics* **2020**, *3*, 672.

[30] M. Hirschberger, L. Spitz, T. Nomoto, T. Kurumaji, S. Gao, J. Masell, T. Nakajima, A. Kikkawa, Y. Yamasaki, H. Sagayama, H. Nakao, Y. Taguchi, R. Arita, T. H. Arima, and Y. Tokura, *Phys. Rev. Lett.* **2020**, *125*, 076602.

[31] D. M. Crum, M. Bouhassoune, J. Bouaziz, B. Schweflinghaus, S. Blugel, and S. Lounis, *Nat. Commun.* **2015**, *6*, 8541.

[32] C. Hanneken, F. Otte, A. Kubetzka, B. Dupe, N. Romming, K. von Bergmann, R. Wiesendanger, and S. Heinze, *Nat. Nanotechnol.* **2015**, *10*, 1039.

[33] A. Kubetzka, C. Hanneken, R. Wiesendanger, and K. von Bergmann, *Phys. Rev. B* **2017**, *95*, 104433.

[34] T. McGuire and R. Potter, *IEEE Trans. Magn.* **1975**, *11*, 1018.

[35] J. Ye, W. He, Q. Wu, H. L. Liu, X. Q. Zhang, Z. Y. Chen, and Z. H. Cheng, *Sci Rep* **2013**, *3*, 2148.

[36] Z. Wang, X. Wang, M. Li, Y. Gao, Z. Hu, T. Nan, X. Liang, H. Chen, J. Yang, S. Cash, and N. X. Sun, *Adv. Mater.* **2016**, *28*, 9370.

[37] M. M. Miller, G. A. Prinz, S. F. Cheng, and S. Bounnak, *Appl. Phys. Lett.* **2002**, *81*, 2211.

[38] J. Tang, L. Kong, W. Wang, H. Du, and M. Tian, *Chin. Phys. B.* **2019**, *28*, 087503.

[39] X. Z. Yu, Y. Onose, N. Kanazawa, J. H. Park, J. H. Han, Y. Matsui, N. Nagaosa, and Y. Tokura, *Nature* **2010**, *465*, 901.

[40] A. K. Nayak, V. Kumar, T. Ma, P. Werner, E. Pippel, R. Sahoo, F. Damay, U. K. Rossler, C. Felser, and S. S. P. Parkin, *Nature* **2017**, *548*, 561.

[41] M. Heigl, S. Koraltan, M. Vanatka, R. Kraft, C. Abert, C. Vogler, A. Semisalova, P. Che, A. Ullrich, T. Schmidt, J. Hintermayr, D. Grundler, M. Farle, M. Urbanek, D. Suess, and M. Albrecht, *Nat. Commun.* **2021**, *12*, 2611.

[42] J. Tang, Y. Wu, W. Wang, L. Kong, B. Lv, W. Wei, J. Zang, M. Tian, and H. Du, *Nat. Nanotechnol.* **2021**, *16*, 1086.

[43] F. Zheng, F. N. Rybakov, A. B. Borisov, D. Song, S. Wang, Z. A. Li, H. Du, N. S. Kiselev, J. Caron, A. Kovacs, M. Tian, Y. Zhang, S. Blugel, and R. E. Dunin-Borkowski, *Nat. Nanotechnol.* **2018**, *13*, 451.

[44] J. Tang, Y. Wu, L. Kong, W. Wang, Y. Chen, Y. Wang, Y. Soh, Y. Xiong, M. Tian, and H. Du, *Natl. Sci. Rev.* **2021**, *8*, nwaa200.

[45] J. Tang, L. Kong, Y. Wu, W. Wang, Y. Chen, Y. Wang, J. Li, Y. Soh, Y. Xiong, M. Tian, and H. Du, *ACS Nano* **2020**, *14*, 10986.

[46] W. Wei, J. Tang, Y. Wu, Y. Wang, J. Jiang, J. Li, Y. Soh, Y. Xiong, M. Tian, and H. Du, *Adv. Mater.* **2021**, *33*, 2101610.

[47] M. Altthaler, E. Lysne, E. Roede, L. Prodan, V. Tsurkan, M. A. Kassem, H. Nakamura, S. Krohns, I. Kézsmárki, and D. Meier, *Physical Review Research* **2021**, *3*, 043191.

[48] Z. Hou, Q. Zhang, X. Zhang, G. Xu, J. Xia, B. Ding, H. Li, S. Zhang, N. M. Batra, P. Costa, E. Liu, G. Wu, M. Ezawa, X. Liu, Y. Zhou, X. Zhang, and W. Wang, *Adv. Mater.* **2020**, *32*, 1904815.

[49] B. Wang, P. K. Wu, N. Bagues Salguero, Q. Zheng, J. Yan, M. Randeria, and D. W. McComb, *ACS Nano* **2021**, *15*, 13495.

[50] N. Kumar, Y. Soh, Y. Wang, and Y. Xiong, *Phys. Rev. B* **2019**, *100*, 214420.

[51] K. Heritage, B. Bryant, L. A. Fenner, A. S. Wills, G. Aeppli, and Y. A. Soh, *Adv. Funct. Mater.* **2020**, *30*, 1909163.

[52] G. L. Caer, B. Malaman, and B. Roques, *J. Phys. F: Met. Phys.* **1978**, *8*, 323.

[53] I. Lyalin, S. Cheng, and R. K. Kawakami, *Nano Lett.* **2021**, *21*, 6975.

[54] Y. Chen, B. Lv, Y. Wu, Q. Hu, J. Li, Y. Wang, Y. Xiong, J. Gao, J. Tang, M. Tian, and H. Du, *Phys. Rev. B* **2021**, *103*, 214435.

[55] J. Jiang, Y. Wu, L. Kong, Y. Wang, J. Li, Y. Xiong, and J. Tang, *Acta Mater.* **2021**, *215*, 117084.

[56] Y. Wu, L. Kong, Y. Wang, J. Li, Y. Xiong, and J. Tang, *Appl. Phys. Lett.* **2021**, *118*, 122406.

[57] Y. Wu, J. Tang, B. Lyu, L. Kong, Y. Wang, J. Li, Y. Soh, Y. Xiong, M. Tian, and H. Du, *Appl. Phys. Lett.* **2021**, *119*, 012402.

[58] Q. Du, M. G. Han, Y. Liu, W. Ren, Y. Zhu, and C. Petrovic, *Advanced Quantum Technologies* **2020**, *3*, 2000058.

[59] K. Shibata, X. Z. Yu, T. Hara, D. Morikawa, N. Kanazawa, K. Kimoto, S. Ishiwata, Y. Matsui, and Y. Tokura, *Nat. Nanotechnol.* **2013**, *8*, 723.

[60] J. H. Scofield, *American Journal of Physics* **1994**, *62*, 129.

[61] A. Vansteenkiste, J. Leliaert, M. Dvornik, M. Helsen, F. Garcia-Sanchez, and B. Van Waeyenberge, *AIP Advances* **2014**, *4*, 107133.

# Supporting Information

## Combined Magnetic Imaging and Anisotropic Magnetoresistance Detection of Dipolar Skyrmions

Jin Tang[1,2], Jialiang Jiang[2], Ning Wang[2], Yaodong Wu[2], Yihao Wang[2], Junbo Li[2], Y. Soh[3], Yimin Xiong[1,2], Lingyao Kong[1]*, Shouguo Wang[4], Mingliang Tian[1,2], and Haifeng Du[2]*

[1]School of Physics and Optoelectronics Engineering Science, Anhui University, Hefei, 230601, China

[2]Anhui Province Key Laboratory of Condensed Matter Physics at Extreme Conditions, High Magnetic Field Laboratory, HFIPS, Anhui, Chinese Academy of Sciences, Hefei, 230031, China

[3]Paul Scherrer Institute, 5232, Villigen, Switzerland

[4]School of Materials Science and Engineering, Anhui University, Hefei 230601, China

Keywords: single skyrmion chain, anisotropic magnetoresistance, magnetic imaging

*email: LyKong@ahu.edu.cn; duhf@hmfl.ac.cn

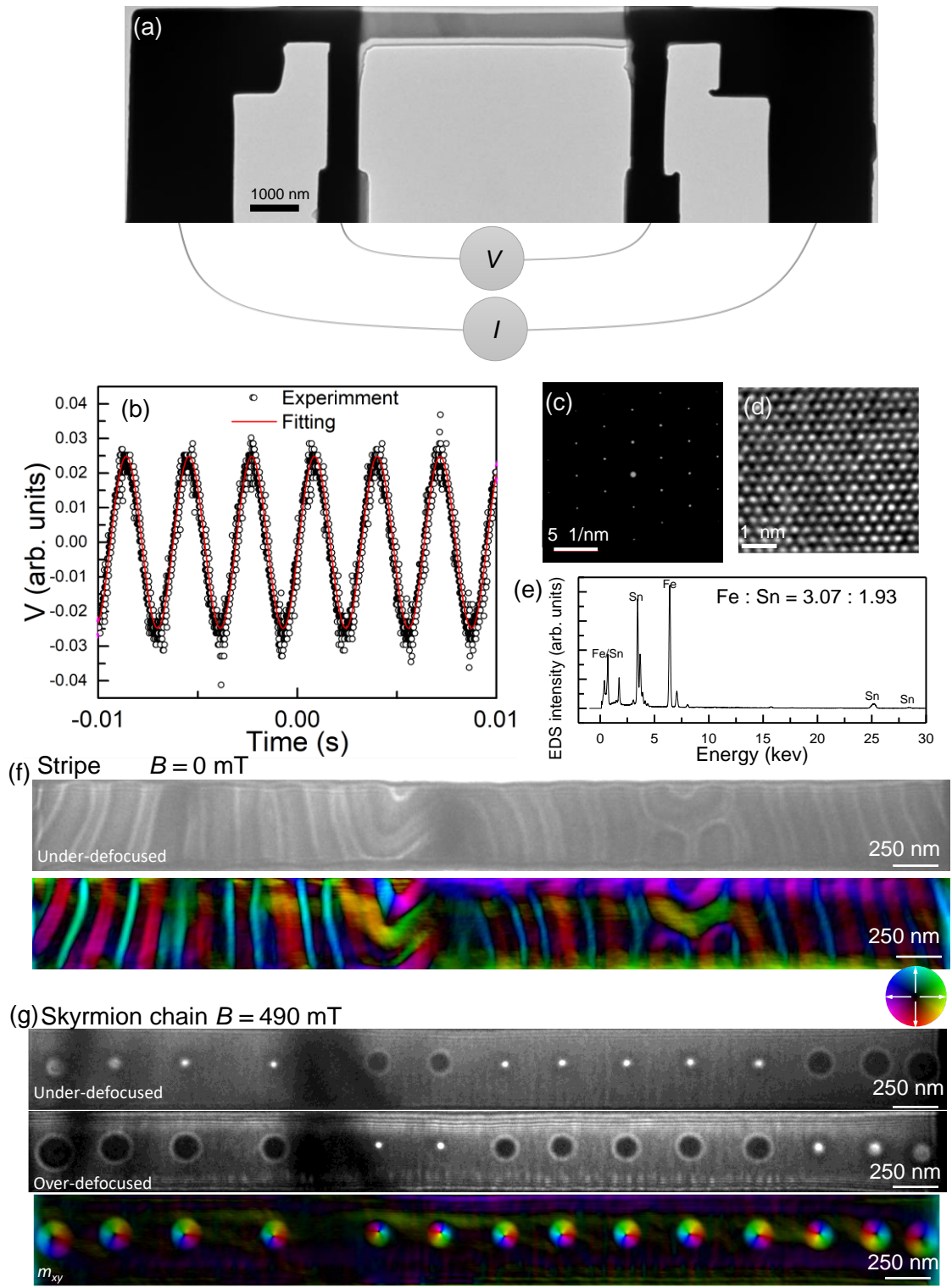

**Figure S1.** a) Experimental setup for the combined magnetic imaging and AMR detection of a single skyrmion chain in the Fe$_3$Sn$_2$ nanostripe. b) Measured voltage profile $V$ between the two ends of the nanostripe when applying a sine wave of input current $I$ with the lock-in technique. The frequency is set at 317 Hz. c)-e) Electronic diffraction pattern (c), high-

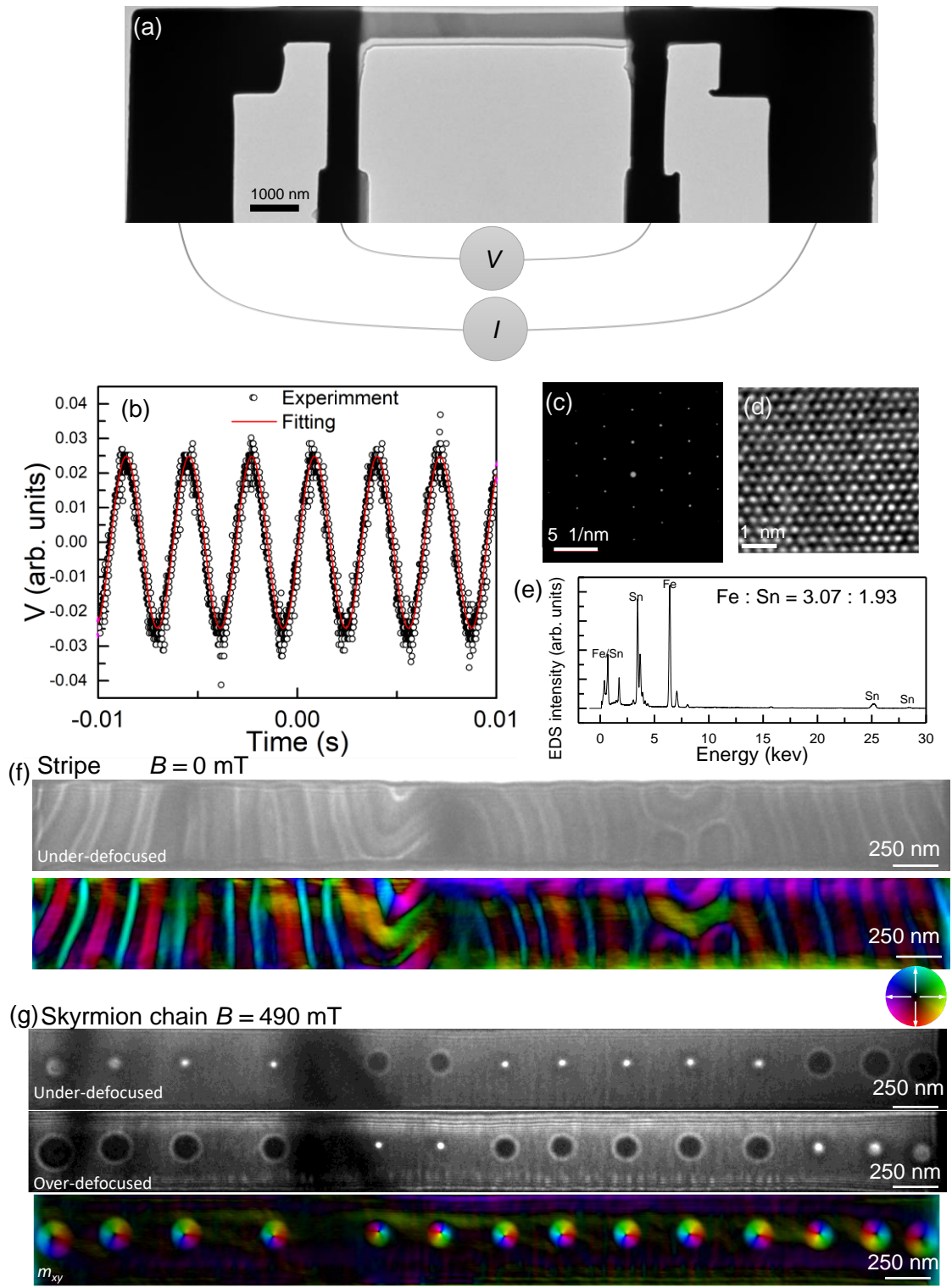

**Figure S1.** a) Experimental setup for the combined magnetic imaging and AMR detection of a single skyrmion chain in the Fe$_3$Sn$_2$ nanostripe. b) Measured voltage profile $V$ between the two ends of the nanostripe when applying a sine wave of input current $I$ with the lock-in technique. The frequency is set at 317 Hz. c)-e) Electronic diffraction pattern (c), high-

resolution scanning transmission electron microscopy (d), and energy dispersion spectrum (EDS) (e) of the $Fe_3Sn_2$(001) device reveal good kagome-lattice order. The atomic ratio of Fe to Sn determined from EDS is 3.07:1.93. f) Defocused Fresnel imaging and retrieved in-plane magnetization mapping of stripe domains at $B = 0$ mT. g) Defocused Fresnel images and retrieved in-plane magnetization mapping of a single skyrmion chain at $B = 490$ mT. The colors represent the in-plane magnetization amplitude and orientation based on the color wheel. $\varphi_B = 0°$.

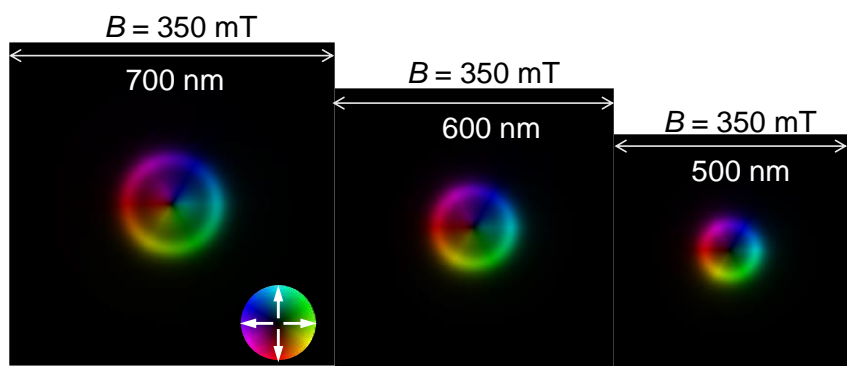

**Figure S2.** Simulated overall in-plane magnetization mapping of single skyrmions at $B =350$ mT in confined geometries with different sizes.

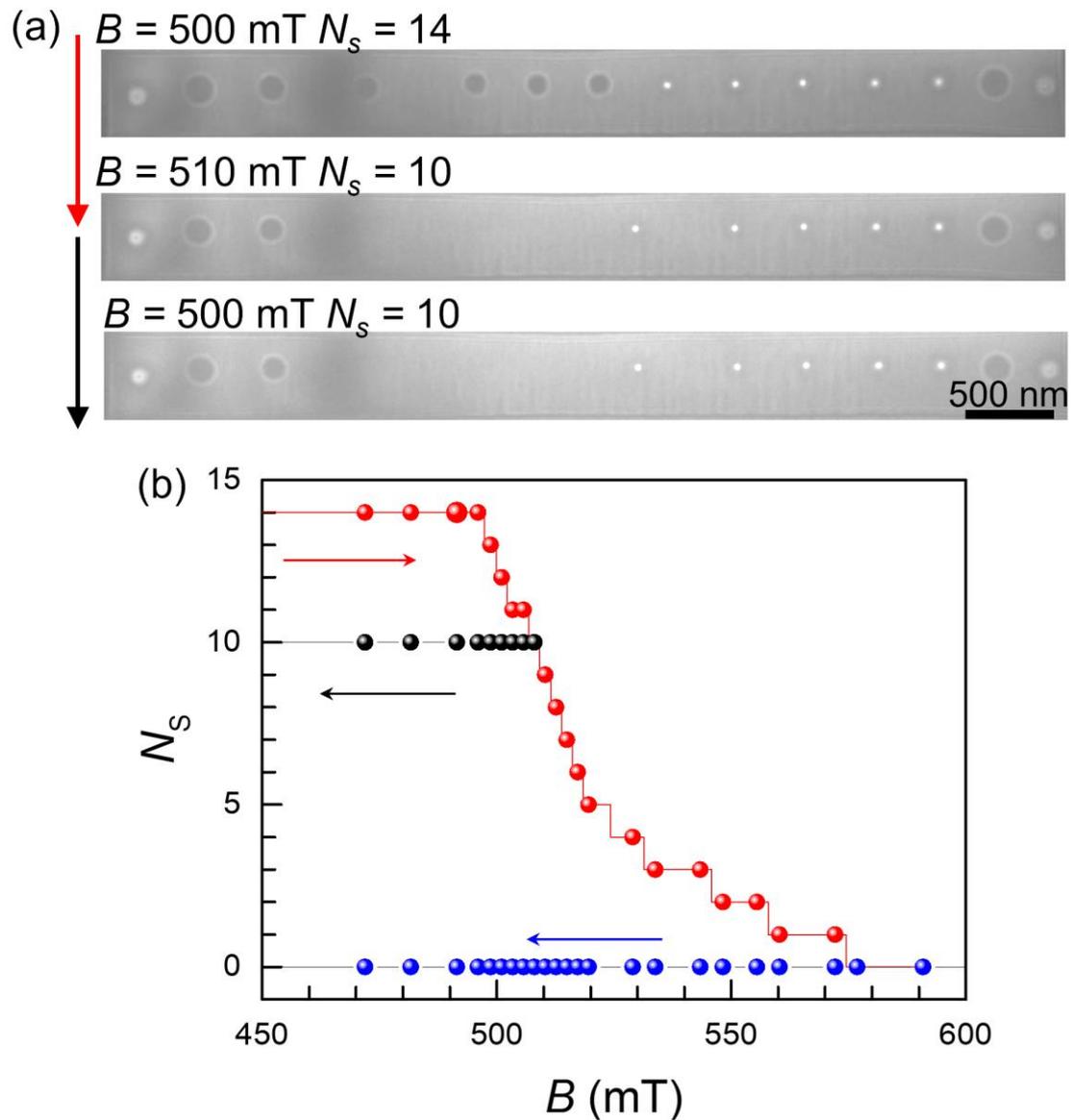

**Figure S3.** a) $N_s = 14$ of a single skyrmion chain turns to 10 when increasing the field to 510 mT and remains 10 when decreasing the field back to 500 mT. b) Experimental skyrmion count $N_s$ during the cycling of magnetic field $B$. $\varphi_B = 0°$.

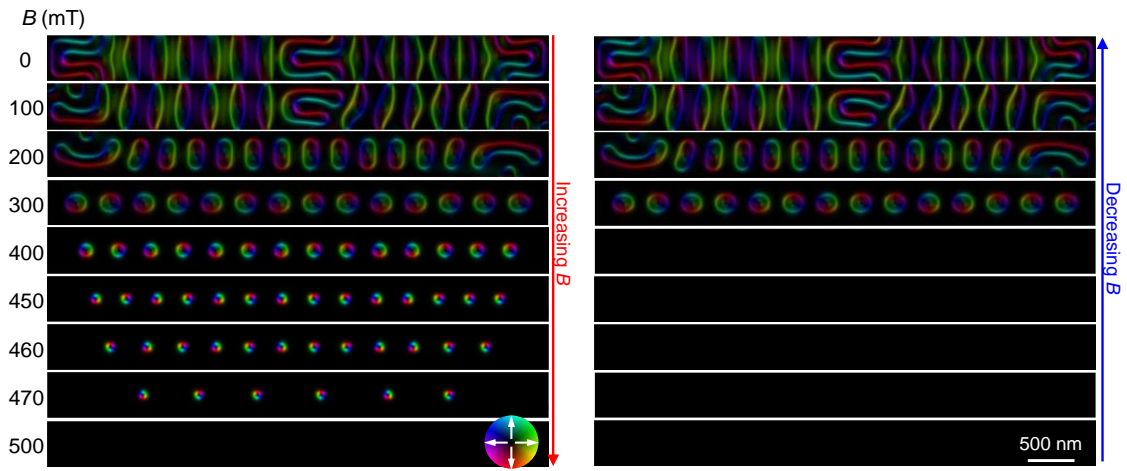

**Figure S4.** Simulated field-driven magnetic evolution between skyrmion chains and stripes in the increasing and decreasing field processes that correspond to the experiments. The colors represent the in-plane magnetization amplitude and orientation based on the color wheel. $\varphi_B = 0°$.

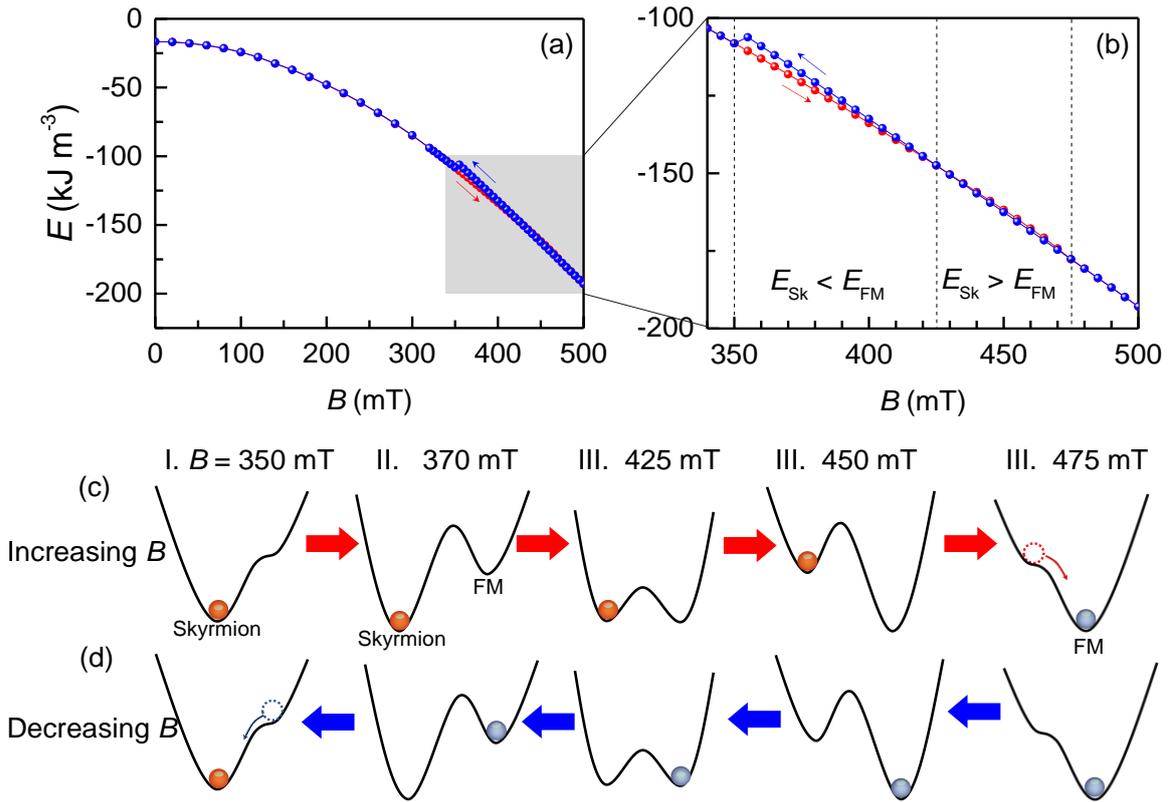

**Figure S5.** Simulated Magnetic field $B$ dependence of total free energy density $E$ during the magnetic evolution between skyrmion chains and stripes. Skyrmion chain and ferromagnet coexist in the field region between 350 and 475 mT. The energy of skyrmion chain $E_{Sk}$ is smaller than that of FM $E_{FM}$ at $B < 425$ mT. The energy of the skyrmion chain $E_{Sk}$ is larger than that of the FM $E_{FM}$ at $B > 425$ mT. a) $0 < B < 500$ mT. b) $345$ mT $< B < 500$ mT. c) Schematic energy profile for transforming from the skyrmion state to FM state in the increasing $B$ process. d) Schematic energy profile for transforming from the FM state to the skyrmion state in the decreasing $B$ process.

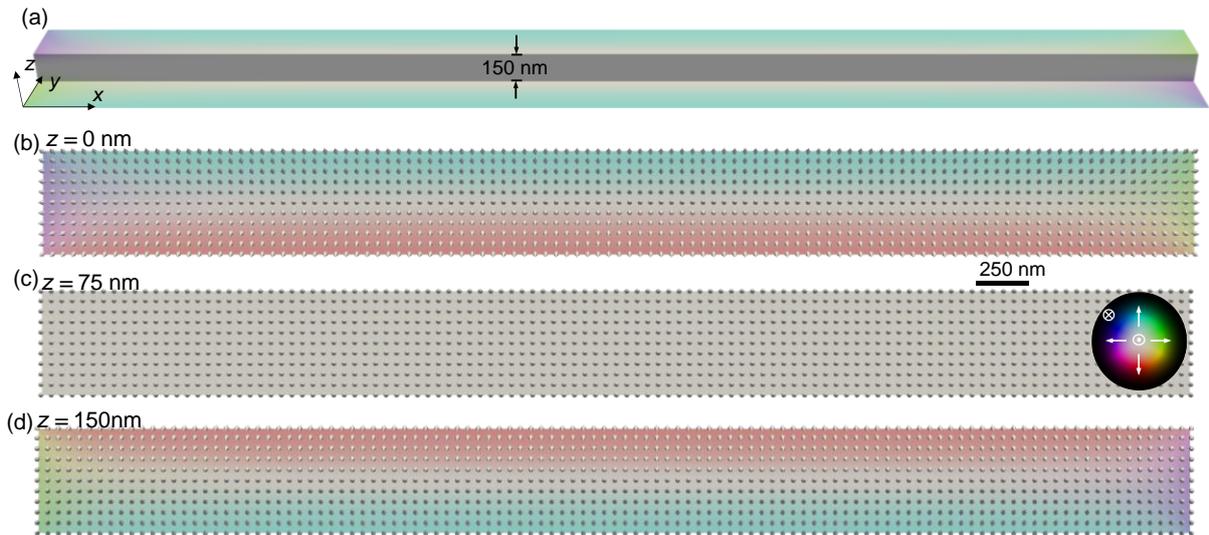

**Figure S6.** Simulated magnetization of a ferromagnet at $B = 400$ mT. a) Overall 3D magnetization. b)-d) Magnetization at the bottom layer $z = 0$ nm (b), the middle layer $z = 75$ nm (c), and the top layer $z = 150$ nm (d). The colors represent the magnetization based on the color wheel. White and black denote the magnetization is out-of-plane in the up and down orientation, respectively. $\varphi_B = 0°$.

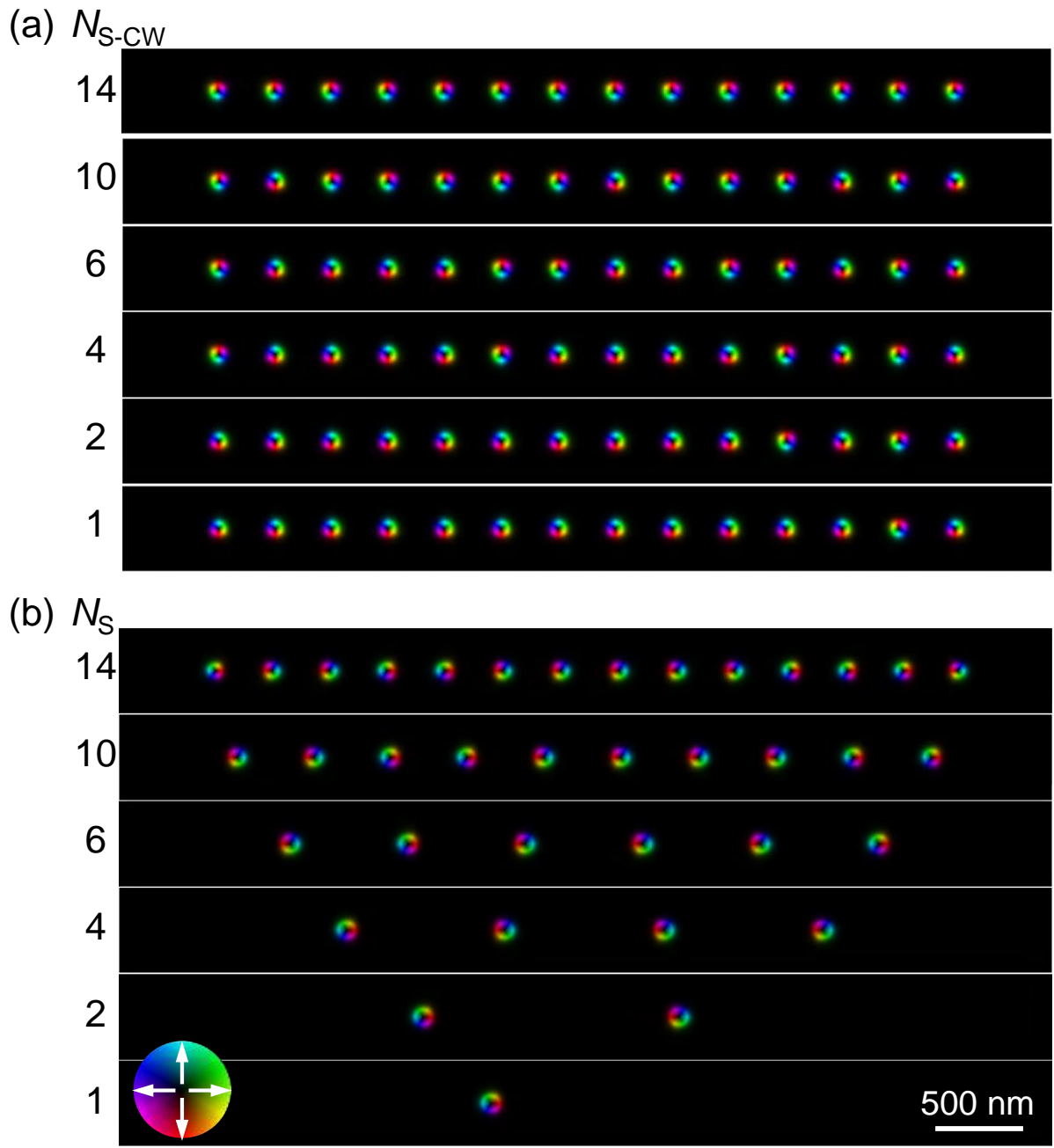

**Figure S7.** a) Simulated overall in-plane magnetization mapping of a single skyrmion chain at $B = 455$ mT. The total skyrmion count $N_s$ is fixed at 14. We varied the count of skyrmion with clockwise rotation $N_{s\text{-cw}}$. b) Simulated overall in-plane magnetization mapping of a single skyrmion chain with varied skyrmion count $N_s$ at $B = 455$ mT. The colors represent the in-plane magnetization amplitude and orientation based on the color wheel. $\varphi_B = 0°$.

8